\documentclass[11pt,a4paper]{article}

\usepackage[paperwidth=24cm]{geometry}
\usepackage[T1]{fontenc}
\usepackage[latin1]{inputenc}
\usepackage{graphicx}
\usepackage{amsmath}
\usepackage{amssymb}

\newcommand{\bl}{\boldsymbol}

\newcommand{\lef}{\left(\,}
\newcommand{\rig}{\,\right)}
\newcommand{\ph}{\phantom}

\newcommand{\eq}{\,=\,}
\newcommand{\ma}{\,+\,}
\newcommand{\me}{\,-\,}
\newcommand{\imp}{\quad \Rightarrow \quad}
\begin{document}

\title{\textbf{Killing-Yano Tensors of Order $n-1$}}

\author{\textbf{Carlos  Batista}\\
\small{Departamento de F\'{\i}sica}\\
\small{Universidade Federal de Pernambuco}\\
\small{50670-901 Recife-PE, Brazil}\\
\small{carlosbatistas@df.ufpe.br}}
\date{}





\maketitle
\begin{abstract}
The properties of a Killing-Yano tensor of order $n-1$ in an $n$-dimensional manifold are investigated. The integrability conditions are worked out and all metrics admitting a Killing-Yano tensor of order $n-1$ are found. It is pointed out a connection between such tensors and a generalization of the concept of angular momentum. A theorem on how to generate closed conformal Killing vectors using the symmetries of a manifold is proved and used to find all Killing-Yano tensors of order $n-1$ of a maximally symmetric space. \textsl{(Keywords: Killing-Yano tensors, Killing tensors, Conformal Killing vectors, General relativity, Angular momentum, Maximally symmetric spaces)}
\end{abstract}

\section{Introduction}

It is hard to overemphasize the relevance of conservation laws to the development of Physics and Mathematics. Indeed, the existence conserved quantities are of fundamental importance in the integration of equations of motion and differential equations in general, specially when these equations are non-linear. Particularly, in General Relativity it is well-known that the Killing vectors can be used to construct scalars that are conserved along geodesic curves. For instance, the $4$-dimensional Schwarzschild spacetime admits four independent Killing vectors, so that the geodesics in this manifold can be found without solving the non-linear geodesic equation. Nevertheless, in the case of the $4$-dimensional Kerr spacetime the issue of finding the geodesic trajectories is much more complicated, since in this case only two Killing vectors are available. However, in 1968 Carter was able to find a new conserved scalar along Kerr geodesics that enabled him to fully integrate the geodesic equation in such spacetime \cite{Carter-constant}. Differently from the conserved scalars associated to Killing vectors, which are linear on the tangent vector of the geodesic, the constant found by Carter is quadratic. Later, Walker and Penrose proved that Carter's constant is a consequence of the fact that Kerr metric admits a Killing tensor of order two \cite{Walk-Pen}. Killing tensors are symmetric generalizations of the concept of Killing vectors that also lead to conservation laws along geodesic curves, so that they are valuable tools in  General Relativity. Indeed, the Killing tensor of Kerr spacetime was the key ingredient to allow the integration of Klein-Gordon and Dirac equations \cite{Carter-KleinG,Chandra-Dirac}, as well as the separability of electromagnetic and gravitational perturbations in Kerr background \cite{Teukolsky}.

Besides the symmetric Killing tensors, there are also anti-symmetric generalizations of the Killing vectors called Killing-Yano tensors \cite{Yano,Tachibana-KY}. These tensors do also lead to conserved quantities along geodesics. In addition, Killing-Yano tensors can be used to construct Killing tensors, since the square of a Killing-Yano is a Killing tensor of order two. In this sense, Killing-Yano tensors can be seen as more fundamental objects than the Killing tensors. In particular, it turns out that the Killing tensor of the Kerr spacetime is just the square of a Killing-Yano tensor \cite{Collinson,Steph_KY}, so that instead of working with the Killing tensor, in this case we can deal with the Killing-Yano tensor with advantages. Recently, Killing-Yano tensors proved to be valuable tools in higher-dimensional spacetimes as well. It was proved by  Frolov and Kubiz\v{n}\'{a}k that the whole family of Kerr-NUT-(A)dS spacetimes, in arbitrary dimension, admits a collection of Killing-Yano tensors that generate conserved quantities just enough to enable the explicit integration of the geodesic equation \cite{Frolov_KY,Kubiz}. These Killing-Yano tensors were also used to separate the Klein-Gordon and Dirac equations in Kerr-NUT-(A)dS background \cite{Frol-KG,Oota}.

In addition to General Relativity, Killing-Yano tensors are also of relevance in other branches of Physics. For instance, in conservative classical systems the solutions of the equations of motion can be seen as geodesics of certain curved spaces, so that Killing-Yano tensors can be used to find the trajectories of these systems \cite{KY-ClassMech}. Such tensors have also been used to construct symmetry operators commuting with the D'Alembertian and Dirac operators \cite{Benn-DiracSymme}, and to solve Maxwell's equation in curved spaces \cite{DebyePot}.


In the present article it will be investigated the Killing-Yano tensors of order $n-1$ in manifolds of dimension $n$. In Sec. \ref{Sec-KY}, Killing-Yano tensors and its associated conserved charges are reviewed. Then, the particular features of the Killing-Yano tensors of order $n-1$ are investigated in Sec. \ref{Sec.KYn-1}. First, it is shown that the existence of such a tensor is equivalent to the existence of a closed conformal Killing vector and then the integrability conditions of the latter object are analysed. After this, the metrics of manifolds admitting a Killing-Yano tensor of order $n-1$ are explicitly found. Then, in Sec. \ref{SubSec.Ang.Mom} it is shown how such tensors can be associated to a generalized angular momentum. In addition, in Sec. \ref{SubSec.Theorem} it is proved a useful theorem stating that one can use the Killing vectors of a space to generate Killing-Yano tensors of order $n-1$. Finally, in Sec. \ref{Sec.Examples}, two examples are worked out. In particular, all closed conformal Killing vectors of a maximally symmetric space are found.

\section{Killing-Yano Tensors}\label{Sec-KY}
In what follows we shall deal with a Riemannian manifold $(M,\bl{g})$ of arbitrary signature and arbitrary dimension $n$. The manifold is assumed to be endowed with the Levi-Civita connection, $\nabla_a$, which is compatible with the metric and torsion-free. The formal distinction between $1$-forms and vector fields will be often ignored in what follows, as the metric provides a one-to-one map between these objects. In this context, a vector field $\bl{K}$ is called a Killing vector whenever it obeys to the so called Killing equation:
$$ \nabla_a\, K_b \,+\, \nabla_b\, K_a \,=\, 0 \,. $$
The concept of Killing vectors can be easily generalized, giving rise to what are called the Killing tensors. A tensor $\bl{K}$ is called a Killing tensor of order $p$ if it is totally symmetric and obeys to the following equation:
\begin{equation}\label{Killing Tensor}
  \nabla_{(a}\, K_{b_1b_2\cdots b_p)} \,=\, 0 \,,
\end{equation}
where the round brackets means that the enclosed indices are symmetrized. In particular, the metric is a Killing tensor of order two. The Killing vectors and tensors are important because they lead to conserved scalars along the geodesic motion. Indeed, let $\bl{T}$ be an affinely parameterized geodesic vector field, $T^a\nabla_aT^b=0$. Then, it easily follows from (\ref{Killing Tensor}) that the scalar
$$ C \,=\, K_{a_1a_2\cdots a_p}\, T^{a_1}T^{a_2}\cdots T^{a_p}  $$
is conserved along the geodesic curve tangent to $\bl{T}$, \textit{i.e.},  $T^a\nabla_a C =0$. Note that a Killing tensor of order $p$ provides a conserved scalar that is of degree $p$ in the velocity $\bl{T}$. In particular, a Killing vector lead to a conserved quantity that is linear in the velocity, which can be interpreted as the momentum of the motion in the direction of the Killing vector.

Instead of symmetric tensors, the idea of Killing vectors can also be generalized as skew-symmetric tensors \cite{Yano}. A totally skew-symmetric tensor $\bl{Y}$ is called a Killing-Yano tensor of order $p$ if it obeys to the following generalization of the Killing equation:
\begin{equation}\label{KY-equation}
  \nabla_{a}\,Y_{b_1b_2\ldots b_p} \,+\, \nabla_{b_1}\,Y_{ab_2\ldots b_p}\,=\, 0\,.
\end{equation}
Note that the above equation is tantamount to the following one:
\begin{equation}\label{KY-derivative}
  \nabla_{a}\,Y_{b_1b_2\ldots b_p} \,=\, \nabla_{[a}\,Y_{b_1b_2\ldots b_p]}\,,
\end{equation}
where the square brackets means that the indices on the right hand side are completely anti-symmetrized. Killing-Yano tensors are also associated to conservation laws. Indeed, let $\bl{T}$ be an affinely parameterized geodesic vector field. Then, we can use the Killing-Yano tensor $\bl{Y}$ to define the tensor
\begin{equation}\label{Conserved-P}
  P_{b_2\ldots b_p} \,=\, T^{b_1}\,Y_{b_1b_2\ldots b_p}
\end{equation}
that is conserved along the geodesic generated by $\bl{T}$. In order to see that such tensor is constant along the geodesic, note that
$$  T^{a}\nabla_a\,P_{c_2\ldots c_p}  \,=\, T^aT^{b}\,\nabla_aY_{bc_2\ldots c_p} \,=\, T^aT^{b}\,\frac{1}{2}\lef\nabla_{a}Y_{bc_2\ldots c_p} \,+\, \nabla_{b}Y_{ac_2\ldots c_p}\rig \,=\, 0\,. $$
In particular, since $\bl{P}$ is conserved it follows that the scalar constructed using its square, $P^2 \equiv  P_{b_2\ldots b_p} P^{b_2\ldots b_p} $, is also conserved. Note that such scalar can be written as
\begin{equation}\label{Conserved scalar}
  P^2 \,=\, K_{ab}\, T^{a}T^{b} \;,\; \textrm{ where } \,  K_{ab} = Y_{a}^{\ph{a}c_2\cdots c_p}\,Y_{bc_2\cdots c_p} \,.
\end{equation}
Since the conserved scalar $P^2$ is quadratic in the velocity $\bl{T}$, it follows that the tensor $K_{ab}$ defined in (\ref{Conserved scalar}) must be a Killing tensor. Therefore, we can take the square of a Killing-Yano tensor to construct a Killing tensor of order two. Generally, if $\bl{Y}$ and $\bl{Z}$ are both Killing-Yano tensors of order $p$, then the tensor
\begin{equation}\label{KillingTensorKY}
  Q_{ab} \,=\, Y_{(a}^{\ph{a}c_2\cdots c_p}\, Z_{b)c_2\cdots c_p}
\end{equation}
is a Killing tensor of order two. However, there are Killing tensors that are not the square of Killing-Yano tensors. The later issue was investigated by Collinson \cite{Collinson} and Stephani \cite{Steph_KY}, where it was proved that in vacuum 4-dimensional spacetimes of Petrov type D every Killing tensor turns out to be the square of a Killing-Yano tensor of order two, see also \cite{Ferrando}. More about Killing tensors and conformal Killing tensors of order two is available in \cite{Coll}. Regarding Killing-Yano tensors of order two, some interesting results in 4-dimensional spacetimes can be found in Refs. \cite{Dietz} and \cite{Hall-KY}.

One can say that any scalar function is a Killing-Yano tensor of order zero, while a Killing-Yano tensor of order one is a Killing vector. As any skew-symmetric tensor of rank $n$, a Killing-Yano tensor of order $n$ must be a multiple of the volume-form of the manifold,
\begin{equation}\label{KY-order n}
  Y_{a_1a_2\ldots a_n} \,=\, \lambda\, \epsilon_{a_1a_2\ldots a_n}\,.
\end{equation}
Then, using (\ref{KY-derivative}) and the fact that $\bl{\epsilon}$ is covariantly constant, we conclude that $\lambda$ must be a constant. The Killing tensor associated to (\ref{KY-order n}) is just the metric, apart from a meaningless constant multiplicative factor. Therefore, we can say that the Killing-Yano tensors of order $p=0$, $p=1$ and $p=n$ are, in a sense, trivial. The simplest non-trivial case is $p=n-1$. For instance, in \cite{Popa} it was shown an example in which Killing-Yano tensors of order $p=3$ in $n=4$ dimensions lead to non-obvious conserved quantities. In spite of the simplicity, there are very few comments about Killing-Yano tensors of order $n-1$ in the literature and the intent of the present article is to fill this gap. However, it is worth noting that some illuminating comments on the case of Killing-Yano tensors of order $3$ in $4$-dimensional Lorentzian manifolds can be found in \cite{Dietz}.

\section{The Killing-Yano Tensors of Order $n-1$}\label{Sec.KYn-1}

Let $\bl{Y}$ be a Killing-Yano tensor of order $n-1$. Then, since it is a totally anti-symmetric tensor, it can be seen as a differential form. In particular, $\bl{Y}$ is Hodge dual of some 1-form $\bl{\xi}$:
\begin{equation}\label{Dual Y}
  \bl{Y} \,=\, \star \,\bl{\xi} \quad \Rightarrow\quad  Y_{a_2a_3\cdots a_n} \,=\, \xi^{a_1}\,\epsilon_{a_1a_2a_3\cdots a_n}\,.
\end{equation}
Inserting the latter relation into (\ref{KY-derivative}) and using the fact that every $n$-form must be proportional to the volume form lead us to
$$ \epsilon_{b_1b_2b_3\cdots b_n}\,\nabla_a\,\xi^{b_1} \eq \nabla_{[a}Y_{b_2b_3\cdots b_n]} \eq S \, \epsilon_{ab_2\cdots b_n}\,,  $$
where $S$ is some scalar function. Then, contracting the above equation with $\epsilon^{cb_2\cdots b_n}$ we find that $\nabla_a\xi_c = h\,g_{ac}$, where $h$ is some function. Taking the trace of this equation, we conclude that $h=\frac{1}{n}\nabla_a\xi^a$. Therefore, the Hodge dual of a Killing-Yano tensor of order $n-1$ obeys to the following equation:
\begin{equation}\label{CKV-closed}
  \nabla_a\, \xi_b \eq h\, g_{ab} \;,\;\textrm{ where }\; h \eq \frac{1}{n}\, \nabla_a\xi^a \,.
\end{equation}
The above equation is tantamount to say that $\bl{\xi}$ is a closed conformal Killing vector, since $\nabla_{[a} \xi_{b]}=0$ and $\nabla_{(a} \xi_{b)}\propto g_{ab}$\,. Conversely, suppose that $\bl{\xi}$ is a closed conformal Killing vector, meaning that Eq. (\ref{CKV-closed}) is satisfied, then defining $\bl{Y}$ as in Eq. (\ref{Dual Y}) we find
$$ \nabla_{a}Y_{b_2b_3\cdots b_n} \eq  \epsilon_{b_1b_2b_3\cdots b_n}\, \nabla_{a} \xi^{b_1} \eq h\,\epsilon_{ab_2b_3\cdots b_n} \quad \Rightarrow\quad \nabla_{(a}Y_{b_2)b_3\cdots b_n} \eq 0\,, $$
so that $\bl{Y}$ is a Killing-Yano tensor. Therefore, we conclude that \emph{$\bl{Y}$ is a Killing-Yano tensor of order $n-1$ if, and only if, its Hodge dual is a closed conformal Killing vector}. Then, instead of working with $\bl{Y}$ one can, equivalently, deal with $\bl{\xi}$. Actually, this result is just a particular case of the general statement that a differential form is a Killing-Yano tensor if, and only if, its Hodge dual is a closed conformal Killing-Yano tensor \cite{Frolov_KY}.

Let us now display some geometric properties of a closed conformal Killing vector. Since $\bl{\xi}$ is a closed 1-form, it follows that locally we can find some function $\lambda$ such that $\xi_a= \nabla_a\lambda$. So, as a vector field, $\bl{\xi}$ is orthogonal to the family of hyper-surfaces of constant $\lambda$. Moreover, contracting Eq. (\ref{CKV-closed}) with $\xi^a$ we see that $\bl{\xi}$ is a geodesic vector field, although generally not affinely parameterized. Now, define $f$ to be the squared norm of $\bl{\xi}$,
$$ f \eq \xi^a\, \xi_a \,. $$
Then, taking the derivative of $f$ lead us to
\begin{equation}\label{Xi=Df}
  \nabla_a\,f \eq 2\,\xi^b\,\nabla_a\xi_b \eq 2h\,\xi_a \,.
\end{equation}
Thus, applying $\nabla_c$ on the above identity and then taking the skew-symmetric part of the resulting equation lead us to:
\begin{equation}\label{Dh=Xi}
  \nabla_c\nabla_a\,f \eq 2\,(\nabla_ch) \,\xi_a \ma 2\,h^2\,g_{ca} \quad \Rightarrow \quad \xi_{[a}\nabla_{c]}h \eq 0 \quad \Rightarrow \quad \nabla_ch \,\propto\, \xi_c\,.
\end{equation}

As pointed out in the preceding section, one can use Killing-Yano tensors to construct Killing tensors of order two, but the converse generally is not true. In particular, if  $\bl{Y}=\star\,\bl{\xi}$ is a Killing-Yano tensor of order $n-1$ then using Eqs. (\ref{Conserved scalar}) and (\ref{Dual Y}) we find that the associated Killing tensor, apart from an unimportant constant global factor, is
\begin{equation}\label{KillinTensor-Y}
  K_{ab} \eq \xi_a\,\xi_b \me f\,g_{ab} \,.
\end{equation}
More generally, if $\bl{Y}=\star\,\bl{\xi}$ and $\bl{Z}=\star\,\bl{\chi}$ are both Killing-Yano tensors of order $n-1$ then, besides the Killing tensors of the form (\ref{KillinTensor-Y}), one can construct another Killing tensor of order two contracting $\bl{Y}$ and $\bl{Z}$ as shown in (\ref{KillingTensorKY}), such Killing tensor is
\begin{equation}\label{KillingTensor-Q}
  Q_{ab} \eq \xi_{(a}\,\chi_{b)} \me (\xi^c\,\chi_c)\,g_{ab} \,.
\end{equation}


\subsection{Integrability Conditions}\label{SubSec.Integrability}

Now, let us work out the integrability conditions for a closed conformal Killing vector, which is tantamount to studying the integrability conditions of a Killing-Yano tensor of order $n-1$. Using (\ref{CKV-closed}) and the Ricci identity we find that
\begin{equation}\label{RicciIdent}
 g_{bc}\,\nabla_ah \me g_{ac}\,\nabla_bh \eq (\nabla_a\nabla_b \me \nabla_b\nabla_a )\, \xi_c \eq R_{abc}^{\ph{abc}d}\,\xi_d \,.
\end{equation}
Then, contracting the above equation with $g^{ac}$ yields the following relation:
\begin{equation}\label{Dh}
  \nabla_a h \eq \frac{-1}{n-1} \, R_a^{\ph{a}b}\,\xi_b \,.
\end{equation}
In particular, such relation along with (\ref{Dh=Xi}) implies that $\bl{\xi}$ is an eigenvector of the Ricci tensor,
$$ R_a^{\ph{a}b}\,\xi_b \,\propto \, \xi_a \,. $$
Taking the covariant derivative of (\ref{CKV-closed}) and then using Eq. (\ref{Dh}) lead to
\begin{equation}\label{DDxi}
  \nabla_a\nabla_b\,\xi_c \eq \frac{-1}{n-1}\,g_{bc}\, R_a^{\ph{a}d}\,\xi_d \,.
\end{equation}
Such equation along with (\ref{CKV-closed}) implies that in order to know the vector field $\bl{\xi}$ in the whole manifold we just need to know $\xi_a$ and $\nabla_a\xi^a$ in one point of the manifold, the components of $\bl{\xi}$ in the other points can then be obtained integrating equations (\ref{CKV-closed}) and (\ref{DDxi}). In particular, this implies that the maximum number of independent closed conformal Killing vectors in an $n$-dimensional manifold is $n+1$. This maximum number is attained just in maximally symmetric spaces.

Then, inserting Eq. (\ref{Dh}) into (\ref{RicciIdent}) yields the following integrability condition:
\begin{equation}\label{IntegrabilityCond}
  \left[ \,(n-1)\, R_{abc}^{\ph{abc}d} \ma  g_{bc}\,R_{a}^{\ph{a}d} \me g_{ac}\,R_{b}^{\ph{b}d}    \right] \, \xi_d \eq 0 \,.
\end{equation}
Such identity states that in order for a manifold to admit a non-zero solution to Eq. (\ref{CKV-closed}), the curvature must be constrained. In particular, in the case of an Einstein space, a space in which $R_{ab}=\Lambda g_{ab}$, the above integrability condition is equivalent to:
\begin{equation}\label{Weyl condition}
  C_{abcd}\,\xi^d \eq 0 \quad \textrm{ (Einstein spaces)} \,,
\end{equation}
where $C_{abcd}$ is the Weyl tensor. In 4-dimensional Lorentzian manifolds, condition (\ref{Weyl condition}) implies that either the Weyl tensor vanishes or $\bl{\xi}$ is a null vector field and the Petrov type of the Weyl tensor is $N$.  Moreover, in higher dimensions if $\bl{\xi}$ is null and the signature is Lorentzian the constraint (\ref{Weyl condition}) implies that the algebraic type of the Weyl tensor is special according to the CMPP classification \cite{CMPP}, although not necessarily type $N$ \cite{Bel-Deb.Higher}. In the same vein, such condition constrains the algebraic type of the Weyl tensor according to the classification defined in \cite{art4}. Particularly, using the 6-dimensional spinorial formalism presented in \cite{Spin6D}, it follows that if $\bl{\xi}$ is null then its spinorial representation can be chose to be $\xi^{AB}= \chi_1^{[A}\chi_2^{B]}$, so that the condition (\ref{Weyl condition}) can be proved to be equivalent to the following constrains on the spinorial version of the Weyl tensor:
$$ \Psi^{A3}_{\ph{A3}B1}\eq \Psi^{A3}_{\ph{A3}B2}\eq \Psi^{A4}_{\ph{A4}B1}\eq \Psi^{A4}_{\ph{A4}B2}\eq 0 \quad; \quad \Psi^{A1}_{\ph{A1}B1}\ma \Psi^{A2}_{\ph{A3}B2}\eq 0 \,.  $$
However, it is worth noting that, differently from what happens in 4 dimensions, in higher-dimensional Einstein spacetimes with non-vanishing Weyl tensor a vector field $\bl{\xi}$ obeying to (\ref{Weyl condition}) is not necessarily null.


\subsection{Metrics Admitting a Killing-Yano Tensor of Order $n-1$}\label{SubSec.Metrics}

In this section it will be presented the explicit form of the metrics allowing the existence of a Killing-Yano tensor of order $n-1$, which is equivalent to the existence of a closed conformal Killing vector $\bl{\xi}$. In order to accomplish this, let us separate the analysis in two possibilities, the cases $\bl{\xi}$ null and non-null. Before proceeding, it is worth stressing that the results here are all local and that it is always being assumed that the closed conformal Killing vector is non-trivial, $\bl{\xi}\neq0$.

\subsubsection{The case $\bl{\xi}$ null }

If the norm of $\bl{\xi}$ is identically zero, $f=0$, then Eq. (\ref{Xi=Df}) implies that $h=0$. This, in turn, guarantees that $\bl{\xi}$ is a covariantly constant null vector,
$$  \xi_a\,\xi^a \eq 0 \; \textrm{ and }  \;  \nabla_a\,\xi_b \eq 0\,.  $$
In particular, since $\bl{\xi}$ is constant, the Killing-Yano tensor defined in (\ref{Dual Y}) is also constant. In 4 dimensions, spacetimes admitting a covariantly constant null vector field are called $pp$-wave, they represent gravitational waves with well-defined wave surfaces. In arbitrary dimension, the general form of a Lorentzian metric with a covariantly constant null vector is given by \cite{Coley-cont.null.vector}:
$$ ds^2 \eq 2du\,\left[\,dv + H(u, x)\,du + F_j(u, x )\,dx^j\,\right] \ma \tilde{g}_{ij}(u, x)\,dx^i\,dx^j \,,$$
with the indices $i,j$ running from $1$ to $n-2$. Where in the above coordinates the covariantly constant null vector is given by $\bl{\xi}= \partial_v$ or, equivalently, by $\bl{\xi}= du$. Note that the functions in this line element do not depend on the coordinate $v$, which is a consequence of the fact that $\bl{\xi}$ is a Killing vector.

Spacetimes in which all scalars constructed with the curvature, its derivatives and the metric vanish are called vanishing scalar invariant (VSI). Such manifolds are quite important in quantum theories of gravity, since for them the quantum corrections to Einstein-Hilbert action vanish. There is an important class of VSI spacetimes that admit a covariantly constant null vector and in Ref. \cite{Coley-VSI} it was found the general metric of such spacetimes. Particularly, it was proved that, in the CMPP classification \cite{CMPP}, the possible algebraic types of the Weyl tensor in the latter spaces are $III$, $N$ or $O$ \cite{Coley-VSI}.

\subsubsection{The case $\bl{\xi}$ non-null }

Since $\bl{\xi}$ is a closed 1-form, it follows from the Poincar\'{e} lemma that locally we can find some function $\lambda$ such that $\bl{\xi}=d\lambda$. We can then choose such function to be one of the local coordinates. Thus, let us use the coordinates $\{x^j,\lambda\}$, with $i,j\in\{1,2,\cdots,n-1\}$. Moreover, since $d\bl{\xi}\wedge\bl{\xi}=0$, it follows from the Frobenius theorem that the tangent subspaces orthogonal to $\bl{\xi}$ form an integrable foliation. Therefore, the coordinates $\{x^j\}$ can always be chosen in such a way that the coordinate vectors $\partial_j$ are orthogonal to $\bl{\xi}$. In the latter frame, since  $\bl{\xi}$ is assumed to be non-null, we have the following line element:
$$ ds^2 \eq  g_{ij}(\lambda,x)\, dx^i\,dx^j \ma g_{\lambda\lambda}(\lambda,x)\, d\lambda^2  \,.   $$
Since $\bl{\xi}=d\lambda$ we can prove that $g_{\lambda\lambda}= f^{-1}$, where $f=\xi^a\xi_a$. Now, using (\ref{Xi=Df}) in this coordinate frame we have that:
$$  \partial_a f \eq 2\,h \, \xi_a\eq 2\,h \, \delta^{\,\lambda}_a \quad \Rightarrow \quad \partial_jf\eq 0 \imp f\eq f(\lambda) \,. $$
In the same vein, thanks to Eq. (\ref{Dh=Xi}), it follows that the function $h= \frac{1}{n}\nabla_a\xi^a$ depends just on the coordinate $\lambda$. Thus, using this fact into Eq. (\ref{CKV-closed}) lead us to:
$$  h(\lambda)\,g_{ab} \eq \nabla_a\, \xi_b \eq \nabla_a \,\delta^{\,\lambda}_b \eq -\,\Gamma_{ab}^\lambda \eq \frac{1}{2}\, f(\lambda)\lef
\partial_\lambda g_{ab} \me \partial_a g_{b \lambda} \me \partial_bg_{a \lambda}  \rig \,.$$
Such equation is equivalent to the following constraints:
\begin{equation}\label{h=df}
  h(\lambda) \eq  \frac{1}{2}\,\partial_\lambda f \quad;\quad
  \partial_\lambda\, g_{ij} \eq  2\,\frac{h(\lambda)}{f(\lambda)}\, g_{ij}         \,.
\end{equation}
These two constraints, in turn, imply the following:
$$  \partial_\lambda \,g_{ij} = \left[\,\partial_\lambda \ln(f)\,\right]\, g_{ij} \imp g_{ij}(\lambda,x) \eq f(\lambda)\, \tilde{g}_{ij}(x)\,. $$
Thus, we have proved that if a manifold of arbitrary dimension admits a non-null closed conformal Killing vector $\bl{\xi}$, then locally its metric can be put in the following form:
\begin{equation}\label{Metric}
 ds^2 \eq \frac{1}{f(\lambda)}\, (\,d\lambda\,)^2 \ma f(\lambda)\,\tilde{g}_{ij}(x)\, dx^i\,dx^j\,.
\end{equation}
In these coordinates the vector field $\bl{\xi}$ is given by:
\begin{equation}\label{CKV-Standard}
  \bl{\xi} \eq f\,\partial_\lambda  \,\sim\, d\lambda \,.
\end{equation}
Where in the above equation $\bl{\xi}$ was first written as a vector field and then as a 1-form. By means of (\ref{Dual Y}), we find that the Killing-Yano tensor associated to such closed conformal Killing vector is given by:
$$  Y_{a_1a_2\cdots a_{n-1}} \eq \sqrt{|f^n\,\tilde{g}|\,} \, (n-1)! \, \delta_{[a_1}^{\,1}\delta_{a_2}^{\,2}\cdots \delta_{a_{n-1}]}^{\,n-1}
\eq  \sqrt{|f|\,}\, \tilde{\epsilon}_{a_1a_2\cdots a_{n-1}}\,.$$
Where in the above equation $\tilde{g}$ denotes the determinant of $\tilde{g}_{ij}$ and $\tilde{\bl{\epsilon}}$ denotes the volume-form of the submanifold spanned by the coordinates $\{x^i\}$, whose metric is $f\tilde{g}_{ij}$.

\subsubsection{The special case $\bl{\xi}$ non-null and $f$ constant }

A particularly simple case occurs when $f$ is a non-zero constant. In such special case Eq. (\ref{h=df}) implies that $h=0$, so that $\bl{\xi}$ is a covariantly constant non-null vector field. Moreover, because of (\ref{Metric}), we conclude that the metric can be chosen to be
$$  ds^2 \eq \pm\, d\lambda^2 \ma \tilde{g}_{ij}(x)\, dx^i\,dx^j\quad ( \textrm{$f$ constant} )\,,  $$
with $\bl{\xi}=\partial_\lambda$ being the covariantly constant vector field. In these coordinates the components of the Riemann tensor are such that
\begin{equation}\label{Riemann-fconst}
 R_{\lambda abc} \eq  0 \quad  \textrm{and} \quad R_{ijkl} \eq \tilde{R}_{ijkl} \,,
\end{equation}
where the indices $a,b,c$ are arbitrary; $i,j,k,l$ run from $1$ to $(n-1)$; and $\tilde{R}_{ijkl}$ is the Riemann tensor associated to the $(n-1)$-dimensional metric $\tilde{g}_{ij}$.

By means of Eq. (\ref{Riemann-fconst}), one can easily see that $g_{ab}$ is a Ricci-flat metric if, and only if, $\tilde{g}_{ij}$ is Ricci-flat. Particularly, in 4 dimensions if $g_{ab}$ is a vacuum solution then $\tilde{g}_{ij}$ is a 3-dimensional Ricci-flat metric. But, since the Weyl tensor in 3 dimensions is zero, it follows that if the Ricci tensor vanishes identically then the Riemann tensor is zero, $\tilde{R}_{ijkl}=0$. This, in turn, implies that the Riemann tensor of $g_{ab}$ is also zero, as can be grasped from (\ref{Riemann-fconst}). Thus, we conclude that the only 4-dimensional manifold with vanishing Ricci tensor admitting a covariantly constant non-null vector field is the flat space.



\subsection{Relation with the Concept of Angular Momentum}\label{SubSec.Ang.Mom}

In Classical Mechanics, if $\bl{x}$ and $\bl{p}$ are respectively the position and the linear momentum of a particle in 3 dimensions, then the angular momentum is given by the vector $\bl{L} = \bl{x}\times\bl{p}$. Nevertheless, since the vectorial product is not defined in higher dimensions, in Special Relativity the angular momentum shall be understood as an anti-symmetric tensor of rank two \cite{Landau_Field},
\begin{equation}\label{Ang-SR}
 L^{ab}=2\,x^{[a}p^{b]}\,.
\end{equation}
Where $x^a$ and $p^a$ are the position and momentum 4-vectors. In General Relativity the momentum of a particle is well defined, it is just the mass times the normalized tangent vector of its world line. But in general curved spaces the notion of position vector has no intrinsic meaning, so that one cannot define angular momentum as in (\ref{Ang-SR}). However, in the present section it will be proved that in spacetimes endowed with a Killing-Yano tensor of order $n-1$ the associated closed conformal Killing vector can play the role of a position vector.

Let $(M,\bl{g})$ be the flat Euclidean space. Then, one can introduce cartesian coordinates $\{x^a\}$ in which the metric components are given by $g_{ab}=\delta_{ab}$ and the covariant derivative is just the partial derivative. Thus, since the Riemann tensor is identically zero in this space, equations (\ref{CKV-closed}) and (\ref{Dh}) imply that $\partial_a\xi_b=h\delta_{ab}$ and $\partial_ah=0$. The general solution to these equations is $\xi_a=h\,x^a + c_a$, where $h$ and $c_a$ are constants. This leads to the conclusion that, in flat Euclidean spaces, a closed conformal Killing vector is, apart from a global multiplicative scale and a constant translation, the position vector. Therefore, in the present section we shall extrapolate such result and interpret the closed conformal Killing vectors as a kind of generalized position vectors, even in curved spaces.

According to General Relativity, free particles, \textit{i.e.}, particles influenced just by the gravity, move along geodesics. Therefore, we expect from the above paragraphs that if $p^a$ is an affinely parameterized geodesic vector field representing the momentum of a free particle of mass $m$, $p^a\nabla_{a}p^b=0$ and $p^ap_a=-m^2$, and $\xi^a$ is a closed conformal Killing vector, then,
$$L^{ab}= 2\,\xi^{[a}p^{b]}$$
might represent a kind of angular momentum. In particular, the tensor $\bl{L}$ should be conserved along the geodesic. Indeed,
$$  p^c\nabla_c\, L_{ab} \eq p^c\lef \nabla_c\xi_a\, p_b \me p_a\,\nabla_c\xi_b \rig \eq  p^c\lef h\,g_{ca}\, p_b \me p_a\,h\,g_{cb} \rig \eq 0 \,.$$
As a consequence, the scalar $L^{ab}L_{ab}= -2\,[m^2f+(\xi^ap_a)^2]$ is also conserved. Thus, in spacetimes admitting a Killing-Yano tensor of order $n-1$ one can define a generalization of the angular momentum of a particle. More comments on the relation between Killing-Yano tensors and generalizations of the angular momentum can be found in \cite{Dietz,Frolov_KY}.


\subsection{Integrating the Equation of a Closed Conformal Killing Vector with the Help of Symmetries}\label{SubSec.Glass-Garf}

In Refs. \cite{Garf-Glass,Garf-GlassKY}, Garfinkle and Glass have introduced a method that helps in the integration of Killing tensor equation and Killing-Yano equation when the manifold admits a non-null hyper-surface orthogonal Killing vector $\bl{\zeta}$. In \cite{Garf-GlassKY}, such method was also applied to the conformal Killing vector equation. The technic amounts to split these differential equations in the direction parallel to $\bl{\zeta}$ and in the orthogonal part, an approach analogous to the one adopted for obtaining the time evolution of a initial data on a Cauchy surface \cite{Coll2}. In the present section we shall apply this method to split Eq. (\ref{CKV-closed}).

Suppose that $(M,\bl{g})$ admits a non-null hyper-surface orthogonal Killing vector $\bl{\zeta}$. Then, the manifold can be foliated by leaves that are orthogonal to this vector field. Each leaf is a submanifold of dimension $(n-1)$. Let us denote the covariant derivative on such submanifold by $D_a$ and the tensor that projects into these leaves by $\hat{g}_{ab}=g_{ab} - \frac{1}{N}\zeta_a\zeta_b$, where $N=\zeta^a\zeta_a$. Then, by these hypotheses, $\bl{\zeta}$ obeys to the following equation:
$$  \nabla_a\,\zeta_b \eq \frac{1}{N}\,N_{[a}\zeta_{b]}\,,\; \textrm{ where }\; N_a\,\equiv\, D_aN\eq \hat{g}_{ab}\nabla^bN\,. $$
We shall decompose $\bl{\xi}$ as
$$  \xi_a \eq A\,\zeta_a \ma B_a \,,\; \textrm{ where }\;  \zeta^a\, B_a \eq 0 \,.$$
Now, let us separate the indices in the equation satisfied by the closed conformal Killing vector, $\nabla_a\xi_b= h\,g_{ab}$, into the parts that are parallel and orthogonal to $\bl{\zeta}$. Such decomposition has three terms: (1) both indices projected into the leaves orthogonal to $\bl{\zeta}$; (2) one index projected in the direction of $\bl{\zeta}$ and the other index in the orthogonal direction; (3) both indices projected along $\bl{\zeta}$. The equations obtained by means of these projections are respectively given by:
\begin{equation*}
   \left\{
  \begin{array}{l}
    D_a\,B_b \eq  h\, \hat{g}_{ab}\\
    \\
     \mathcal{L}_\zeta\,B_a \eq \frac{1}{2}A\, N_a \;;\; D_a\, A \eq  -\frac{A}{2\,N}\, N_a \\
   \\
   \mathcal{L}_\zeta A \eq \frac{N^{\frac{1}{2}(n-1)}}{n-1}\,D_a\lef N^{\frac{1}{2}(1-n)}\, B^a \rig  \,.\\
    \end{array}
\right.
\end{equation*}
Where $\mathcal{L}_\zeta$ denotes the Lie derivative along the vector field $\bl{\zeta}$. Particularly, the first of these equations means that if $\xi_a$ is a closed conformal Killing vector in $(M,\bl{g})$, then $B_a=\hat{g}_{ab}\xi^b$ is a closed conformal Killing vector in the family of submanifolds orthogonal to $\bl{\zeta}$. For examples showing how this kind of decomposition can be helpful see \cite{Garf-Glass,Garf-GlassKY}.


\subsection{Generating Closed Conformal Killing Vectors Using Symmetries}\label{SubSec.Theorem}

In this section it will be proved a theorem stating that one can use a closed conformal Killing vector and the symmetries of the space to generate other closed conformal Killing vectors. Let $\bl{\xi}$ be a closed conformal Killing vector and $\bl{\eta}$ some Killing vector, then we have that
$$ \nabla_a\xi_b \eq h\,g_{ab} \quad;\quad   \nabla_a\eta_b \eq \nabla_{[a}\eta_{b]}  \,. $$
Now, let us define the following gradient vector field:
$$  \chi_a \eq \nabla_a\,(\xi^b\eta_b) \eq h\,\eta_a \ma \xi^b\,\nabla_{[a}\eta_{b]} \,. $$
Then, the covariant derivative of $\bl{\chi}$ is given by
\begin{equation}\label{DX}
  \nabla_c\,\chi_b \eq (\nabla_ch)\,\eta_b \ma \xi^a\nabla_c\nabla_b\eta_a \eq  \frac{-1}{n-1} \, R_c^{\ph{c}a}\xi_a\,\eta_b \ma \xi^a\,R_{abcd} \, \eta^d  \,.
\end{equation}
Where in the last equality it was used Eq. (\ref{Dh}) and the fact that the second derivative of a Killing vector field can be expressed in terms of the Riemann tensor and the Killing vector itself. Now, using the integrability condition (\ref{IntegrabilityCond}) on the right hand side of (\ref{DX}), we arrive at the following equality:
\begin{equation}\label{X-CKV}
  \nabla_c\,\chi_b \eq  -\frac{(\xi^a\,R_{ad}\,\eta^d)}{n-1}\,\, g_{cb} \,.
\end{equation}
So, $\bl{\chi}$ is also a closed conformal Killing vector. Therefore, we have proved the following:\\
\\
\begin{samepage}
\textbf{Theorem -}\, \nopagebreak[4]\emph{If $\bl{\xi}$ is a closed conformal Killing vector and $\bl{\eta}$ is a Killing vector then the vector field $\chi_a = \nabla_a(\xi^c\eta_c)$ is a closed conformal Killing vector.}
\end{samepage}
\\
\\
Once we have found $\bl{\chi}$ we can, in principle, use it and the Killing vector $\bl{\eta}$ to generate another closed conformal Killing vector, $\kappa_a = \nabla_a(\chi^c\eta_c)$, and so on. However, it is worth recalling that the maximum number of independent closed conformal Killing vectors in an $n$-dimensional manifold is $n+1$. Thus, while using the above procedure to generate closed conformal Killing vectors we can find a vector field that is zero or that is a linear combination, with constant coefficients, of the closed conformal Killing vectors already known.

As a last comment, from Eq. (\ref{X-CKV}) we see that if the Ricci tensor vanishes then the covariant derivative of $\bl{\chi}$ is zero. So, we conclude that if a Ricci-flat manifold admits a closed conformal Killing vector and a Killing vector whose inner product between them is not constant, $\nabla_a(\xi^c\eta_c)\neq0$, then the manifold admits a covariantly constant vector field.

\section{Examples}\label{Sec.Examples}

In this section the results obtained so far will be worked out in two examples. It will be explicitly found all closed conformal Killing vectors of a maximally symmetric space and it will be shown that the FLRW spacetime admits a Killing-Yano tensor of order $n-1$.

\subsection{Maximally Symmetric Spaces}

A maximally symmetric space is a manifold possessing the maximum number of isometries, which is $\frac{1}{2}n(n+1)$ in $n$ dimensions. In the case of Euclidean signature, one can find a coordinate system $\{x^a\}$ in which the line element is:
\begin{equation}\label{Metric-MaxSym}
  ds^2 \eq  \frac{1}{(1\ma \kappa\,r^2)^2} \,\left[\, (dx^1)^2\ma(dx^2)^2\ma\cdots\ma(dx^n)^2  \,\right]\,,
\end{equation}
where
$$  r^2\eq \left[\, (x^1)^2\ma(x^2)^2\ma\cdots\ma(x^n)^2  \,\right]\eq x^a\,x^a\,,\;\;\textrm{ and }\;\;  \kappa\,=\, 0,\,\pm1 \,. $$
The case $\kappa=0$ represents the flat-space, which was already treated in Sec. \ref{SubSec.Ang.Mom}. So, in what follows let us just consider the non-trivial cases $\kappa=\pm1$. The metric of maximally symmetric spaces with other signatures, like de Sitter and Anti-de Sitter spacetimes, can be obtained from (\ref{Metric-MaxSym}) by means of analytical continuations of the form $x^a\rightarrow ix^a$. The independent Killing vectors can be chosen to be
\begin{equation}\label{Killing-MS}
  \bl{\eta}^{ij} \eq x^i\,\partial_j \me x^j\,\partial_i \quad; \quad \bl{\eta}^{i} \eq (1-\kappa\,r^2)\partial_i \ma 2\,\kappa\, x^i\,x^a\partial_a \,.
\end{equation}
Where in the above expressions $i$ and $j$ are not tensorial indices, they are just labels running from $1$ to $n$ with $i\neq j$. The vector fields $\bl{\eta}^{ij}$ are obvious symmetries of the metric (\ref{Metric-MaxSym}), since they generate rotations, which keep $r$ invariant. In order to see that $\bl{\eta}^{i}$ are also Killing vectors, one shall compute the Christoffel symbols of such metric,
$$ \Gamma_{ab}^c \eq  \frac{2\kappa}{1\ma \kappa\,r^2}\,\lef x^c\,\delta_{ab} \me x^a\,\delta_{bc}  \me x^b\,\delta_{ac} \rig\,,  $$
and note that
$$ \nabla_a\,\eta^i_b \eq   \frac{8\,\kappa}{(1\ma \kappa\,r^2)^3}\, \delta^{i[a}\,x^{b]} \imp \nabla_{(a}\,\eta^i_{b)} \eq 0\,. $$

Now, let us look for the closed conformal Killing vectors. A natural guess is that there exists some function $F(r^2)$ such that its gradient is a closed conformal Killing vector $\bl{\xi}^0$:
$$   \bl{\xi}^0 \eq d\,[\,F(r^2)\,] \eq F'\,d(r^2) \eq 2\,F'\,x^a\,dx^a  \,,   $$
where the prime denotes the derivative with respect to $r^2$. Indeed, imposing that $\nabla_a\xi^0_b\propto g_{ab}$ lead to
$$   (1\ma \kappa\,r^2)\,F'' \ma 2\,\kappa\,F' \eq 0 \imp F(r^2) \eq \frac{A}{1\ma \kappa\,r^2} \ma B \,,  $$
where $A$ and $B$ are constants. Choosing $A=-1/2\kappa$ we find that
$$  \bl{\xi}^0 \eq x^a\,\partial_a \,. $$
Using the fact that such space is an Einstein manifold, $R_{ab}=\Lambda g_{ab}$ with $\Lambda=4\kappa(n-1)$, one can readily verify that the integrability conditions (\ref{Dh}) and (\ref{IntegrabilityCond}) are satisfied. The other closed conformal Killing vectors can now be easily found with the help of the theorem presented in Sec. \ref{SubSec.Theorem}. Indeed, according to such theorem the 1-forms
$$  \bl{\xi}^i \eq d\,\left[ \xi^0_a\,\eta^{i\,a} \, \right] \eq d\,\left[ \frac{x^i}{1\ma \kappa\,r^2} \right] \,\sim\, (1+\kappa\,r^2)\partial_i \me 2\,\kappa\, x^i\,x^a\partial_a   $$
are closed conformal Killing vectors, where the symbol $\sim$ means that the tensors are equal if  we use the one-to-one relation between 1-forms and vector fields provided by the metric. Since $(n+1)$ is the maximum number of independent closed conformal Killing vectors in an $n$-dimensional manifold, and since $\{\bl{\xi}^0,\bl{\xi}^i\}$ are linearly independent as vector fields, it follows that we have found all closed conformal Killing vectors of a maximally symmetric space. As a consequence, taking the Hodge dual of the 1-forms $\{\bl{\xi}^0,\bl{\xi}^i\}$, one can find all Killing-Yano tensors of order $n-1$. Moreover, using such Killing-Yano tensors along with Eqs. (\ref{KillinTensor-Y}) and (\ref{KillingTensor-Q}) we can construct $\frac{1}{2}(n+1)(n+2)$ independent Killing tensors of order two.


\subsection{FLRW cosmological model}

The so-called FLRW cosmological model amounts to assume that at large scales the universe can be described as a Lorentzian manifold that is foliated by spatial sections that are homogeneous and isotropic, so that the space-like leaves are maximally symmetric submanifolds. In the case of $n-1$ spatial dimensions, the FLRW line element can be written as
\begin{equation}\label{Metric-FLRW}
  ds^2 \eq -dt^2 \ma  \frac{a^2(t)}{(1\ma \kappa\,r^2)^2} \,\left[\, (dx^1)^2\ma(dx^2)^2\ma\cdots\ma(dx^{n-1})^2  \,\right] \,.
\end{equation}
Now, let us see that such metric can be put in the standard form shown in Eq. (\ref{Metric}). Indeed, defining a new time-like coordinate
$$\lambda \,\equiv \, \int\,a(t)\,dt \,,  $$
we have that the line element is given by
$$ ds^2 \eq -\frac{1}{a^2(t(\lambda))}\,d\lambda^2 \ma  \frac{a^2(t(\lambda))}{(1\ma \kappa\,r^2)^2} \,\left[\, (dx^1)^2\ma(dx^2)^2\ma\cdots\ma(dx^{n-1})^2  \,\right] \,. $$
Note that this form of the line element is exactly equal to the one shown in (\ref{Metric}), with $f=-\,a^{2}$. Then, from Eq. (\ref{CKV-Standard}) we conclude that $$ \bl{\xi} \eq d\lambda \eq a(t)\,dt  $$
is a closed conformal Killing vector. Indeed, one can prove that the covariant derivative of this vector field is given by
$$ \nabla_b\,\xi_c \eq - \,\dot{a}\, g_{bc}  \,, $$
where the dot denotes the derivative with respect to the coordinate $t$. From the above equation, we conclude that $h=- \,\dot{a}$, which agrees with Eq. (\ref{Xi=Df}) and the fact that $f=-\,a^{2}$. By means of (\ref{KillinTensor-Y}), we find that the components of the Killing tensor associated to $\bl{\xi}$ are given, in the coordinates $\{t,x^i\}$, by
$$  K_{bc}  \eq a^2(t)\,\left[\, \delta^{\,t}_b \, \delta^{\,t}_c \ma g_{bc} \,\right] \,. $$
Note that the metric (\ref{Metric-FLRW}) is invariant under rotations performed in the spatial leaves spanned by the coordinates $\{x^i\}$, so that the vector fields $\bl{\eta}^{ij} = x^i\,\partial_j - x^j\,\partial_i$ are Killing vectors of the FLRW spacetime. Where the indices $i$ and $j$ are labels for the spatial coordinates that run from 1 to $n-1$. Thus, in principle, one could use these Killing vectors along with the theorem of Sec. \ref{SubSec.Theorem} in order to generate new closed conformal Killing vectors in addition to $\bl{\xi}$. However, since the Killing vectors $\bl{\eta}^{ij}$ are orthogonal to the closed conformal Killing vector $\bl{\xi}=a\,dt$, it follows that in this case the mentioned theorem does not provide a new symmetry.

\section*{Acknowledgments}
I want to thank CAPES (Coordena\c{c}\~{a}o de Aperfei\c{c}oamento de Pessoal de N\'{\i}vel Superior - Brazil) for the financial support.


\end{document}